# The heavy spring at work

**Amelia Carolina Sparavigna**
Dipartimento di Fisica, Politecnico di Torino
Corso Duca degli Abruzzi 24, Torino, Italy

**Abstract**: Springs are used in simple demonstrations for the students of Physics classes to illustrate the Hooke's Law and harmonic oscillations. The spring is usually considered as a light object that does not possess a mass. What happens if the spring is heavy, that is, its mass is not negligible? This paper aims to discuss this problem as plainly as possible.

**Introduction**
To study the harmonic oscillations, we use the simple pendulum described as an idealized body consisting of a point mass, suspended by a light inextensible cord. In laboratory, the students can measure the period of small oscillations and obtain, with rulers and stop-watches, an evaluation of the acceleration of gravity. Another measurement that seems as simple as that previously proposed, is the measurement of the force constant of a spring. For an ideal spring, we assume valid the Hooke's Law. In a static experience, a body is suspended by the spring. At equilibrium, the elastic force of the spring is balancing the weight of the suspended mass. Using Hooke's Law, we obtain the force constant measuring several deformations of the spring, corresponding to different suspended masses. Another way to measure this force constant is by means of a dynamical method, based on the oscillations of suspended bodies and the measurement of their periods.

What happens if the spring has a small, not negligible mass, that is, it is "heavy"? The discussion of this problem is the aim of the paper. The mass suspended by a spring, which has its mass, becomes a part of a more complex system. Let us call *m* the mass uniformly distributed on the spring and *M* the suspended mass. If *M* is oscillating, we observe that during the motion each section of the spring is moving with its velocity different from that of the suspended mass. We have then the motion of body *M* and that of the spring to determine. Let us remember that for a light spring, that is a spring having no mass, this last motion does not exist: the spring is simply connecting the suspended mass *M* to a fixed body, for instance a stand on a table or the ceiling, giving the interaction between them. Generally, the problem of the heavy spring is solved considering an effective mass of the spring, defined as the mass that must be added to the suspended mass to have a more correct prediction of the behaviour of the system. The effective mass of the spring, assumed as independent whether the direction of the spring-mass system is horizontal, vertical or oblique, is 1/3 of the mass of the spring (a detailed discussion on the effective mass of the spring is proposed at Ref.1).

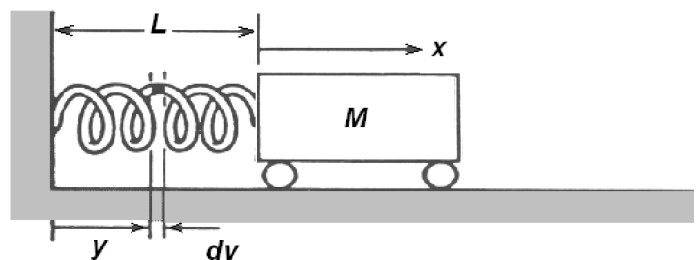

Fig.1. The frame of reference for spring and body.

A simple approach to the effective mass problem is the following. Let us call *m* the mass of the spring, with a free length *L*, and *k* the constant of the spring determined by its stiffness. Take an infinitesimally thin segment of the spring, which is at a distance *y* from the fixed end of the spring (see Fig.1). Its length is *dy* and mass *dm*. To the manner in which the spring is attached, the static deformation of the spring is assumed according to the simple equation: $dy/x = (y/L)$. If the dynamic shape of the spring were the same as the static deformation, the mass and velocity of the spring element *dy* would be:

$$\dot{y} = \frac{dy}{dt} = \left(\frac{y}{L}\right)\frac{dx}{dt} = \left(\frac{y}{L}\right)\dot{x} \quad ; \quad dm = \frac{m}{L}dy \tag{1}$$

The kinetic energy of the spring would be equal to:

$$K = \frac{1}{2}\int_0^L \left[\frac{y}{L}\dot{x}\right]^2 dm = \frac{1}{2}\dot{x}^2 \int_0^L \left(\frac{y}{L}\right)^2 \frac{m}{L}dy = \frac{1}{2}\frac{m\dot{x}^2}{L^3}\int_0^L y^2 dy = \frac{1}{2}\frac{m\dot{x}^2}{L^3}\frac{L^3}{3} = \frac{1}{2}\frac{m}{3}\dot{x}^2 \tag{2}$$

*m*/3 is the effective mass that is often used in evaluating the period of an oscillating mass *M* suspended by a heavy spring: for the period, instead of $\tau=2\pi(M/k)^{1/2}$, it is used $\tau=2\pi((M+m/3)/k)^{1/2}$. This conclusion approximates the true behaviour of a heavy spring. During the oscillation, each part of the spring has its velocity and then it is not described by a linear function of the position.

**The heavy spring at equilibrium**
Since the spring is heavy, when hung vertically from one end, it will stretch under its own weight. We can first find the form of the spring in equilibrium [2]. We assume the heavy spring as an elastic medium with section *S* and length *L*, Young modulus *E* and density $\rho = m/SL$. The correspondence between the Young Modulus and the constant of the spring is given as:

$$\frac{E}{\rho L^2} = \frac{k}{m} \rightarrow E = \frac{k\rho L^2}{m} \tag{3}$$

It is convenient to use as reference position, the spring as it would be if the gravity did not act: that is, vertical and uniform. Let us consider two sections S and S' of the spring. Two points of the spring P on S and Q on S', at the reference positions *x* and *x+Δx*, will be displaced, under the effect of the gravity, by the quantities *y* and *y+Δy* respectively (see Fig.2).

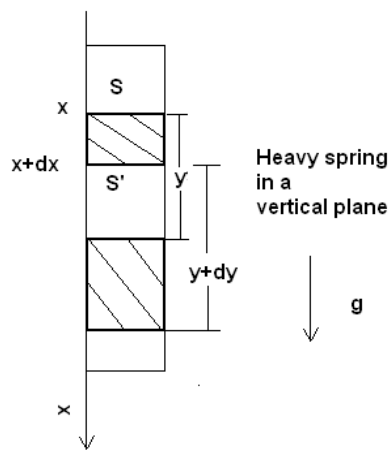

Fig.2 Frame of reference.

The mass element $\Delta m$ from P to Q is subjected to the actions of gravity and of the two tensions $T, T+\Delta T$. Each mass $\Delta m$ is at equilibrium when:

$$\Delta T + \rho S g \Delta x = 0 \tag{4}$$

Since the extension of the element $\Delta y$ is given by the applied force $T+\Delta T$, we have [2]:

$$T + \Delta T = S E \frac{\Delta y}{\Delta x} \tag{5}$$

Processing to the limit as $\Delta x \to 0$, $dT/dx = -\rho S g$, $T = S E\, dy/dx$, we have then:

$$S E \frac{d^2 y}{dx^2} = -\rho S g \tag{6}$$

with the boundary conditions:

$$y = 0 \text{ for } x = 0, \quad \frac{dy}{dx} = 0 \text{ for } x = L \tag{7}$$

That is, the tension vanishes at the lower end [2]. The solution is:

$$y = \frac{\rho g L}{E} x - \frac{\rho g}{2E} x^2 \tag{8}$$

Using Eq.3, we have that:

$$\frac{E}{\rho L^2} = \frac{k}{m} \to E = \frac{k \rho L^2}{m} \tag{9}$$

$$y_o = \frac{mg}{kL} x - \frac{mg}{2kL^2} x^2 = Ax + Bx^2 \tag{10}$$

For $x=L$, we have $y_o = mg/(2k)$. The length of the elongated spring is $L+y_o$.

**The oscillating spring**
In the case that the vertical spring is oscillating about this equilibrium position, the displacement $y$ of each section of the spring is a function of position and time, $y(x, t)$. Each section of the spring satisfies the equation:

$$\frac{\partial^2 y}{\partial x^2} - \frac{\rho}{E} \frac{\partial^2 y}{\partial t^2} + \frac{\rho}{E} g = 0 \tag{11}$$

A simple solution will be: $f(x, t) = \sin(\alpha x) \cos(\omega t + \varphi) + y_o$, where $\alpha$ and $\omega$ are linked by the relation $\alpha = (\rho/E)^{1/2} \omega$. The general solution is a linear combination:

$$f(x, t) = \Sigma_\omega \, A_\omega \sin [\omega (\rho /E)^{1/2} x] \cos (\omega t + \varphi_\omega) + y_o \tag{12}$$

with the sum on all the angular frequencies of the spring.
Imagine that the system is at rest till the initial time, when we let the spring to start its oscillation. The initial velocity of the spring is null: we have then $(\partial y/\partial t)_{t=0}=0$, giving $\varphi_\omega=0$. The general solution of Eq.11 is:

$$y(x, t) = \Sigma_\omega \, A_\omega \sin [\omega (\rho /E)^{1/2} x] \cos (\omega t) + y_o \tag{13}$$

Let us suppose a small displacement $\xi_0$ from the condition of equilibrium, $L+y_o$, that we have under the sole action of gravity. To have this small displacement we need to apply a force to the lower end of the spring. Since at the initial time $t_o=0$, we have the displacement $\xi_o$, the function $y^*(x)=x\xi_o/L$ is a suitable solution for the boundary conditions. At $t_o=0$, the solution for the oscillating case must be equal to function the $y^*(x)$. This condition allows to find coefficients $A_\omega$, imposing that:

$$\Sigma_\omega A_\omega \sin [\omega (\rho /E)^{1/2} x] = y^*(x) = x \, \xi_o / L \tag{14}$$

To complete the discussion of the heavy spring, we have to determine the angular frequency of oscillation. During it, the free end of the spring is subjected to a null tension $(\partial y /\partial x)_{x=L} = 0$, giving then $(\cos[\omega (\rho /E)^{1/2} L]) = 0$, that we can immediately solve as:

$$\omega (\rho /E)^{1/2} L = (n+½)\pi, \quad n = 0,1,2,3... \tag{15}$$

$$\omega = \left(\frac{E}{\rho}\right)^{1/2} \left(n+\frac{1}{2}\right)\frac{\pi}{L} = \sqrt{\frac{k}{m}}\left(n+\frac{1}{2}\right)\pi \tag{15'}$$

Each section of the spring has a motion which is the superposition of many harmonic motions, the respective angular frequencies given by Eq.15'.

**With a mass suspended**
In the case that we have a mass $M$ suspended by the heavy spring, the following boundary conditions in the static case are required:

$$y = 0 \quad for \; x = 0; \quad SE\left(\frac{\partial y}{\partial x}\right)_{x=L} = Mg \quad for \; x = L \tag{16}$$

The static solution is:

$$y = \frac{Mg}{ES}x + \frac{\rho g L}{E}x - \frac{\rho g}{2E}x^2 = \frac{Mg}{kL}x + \frac{gm}{kL}x - \frac{gm}{2kL^2}x^2 = A'x + B'x^2 \tag{17}$$

For $x=L$, we have $y_o=(M+m/2)\,g/k$, which is giving the elongation of the heavy spring with a mass attached. In the case that the mass $M$ is oscillating about this equilibrium position, we have again for the spring the equation:

$$\frac{\partial^2 y}{\partial x^2} - \frac{\rho}{E}\frac{\partial^2 y}{\partial t^2} + \frac{\rho}{E}g = 0 \tag{18}$$

The solution will be $y = f(x, t) + A'x + B'x^2$.

Again, we consider that $f(0,t)$ must be equal to zero, because the section corresponding to $x = 0$ is at a fixed position. A simple solution is: $f(x, t) = \sin(\alpha x) \cos(\omega t + \varphi_\omega)$, where $\alpha$ and $\omega$ are linked by the relations $\alpha = (\rho/E)^{1/2} \omega$. The general solution is a linear combination:

$$y(x, t) = \Sigma_\omega A_\omega \sin[\omega(\rho/E)^{1/2}x] \cos(\omega t) + A'x + B'x^2 \tag{19}$$

where we imagined that at the initial time $t_o$ the system has a null velocity.

As previously done, we imagine that the spring is at equilibrium under the effect of its weight, the weight of the mass $M$ and a force $F$ applied to $M$, maintaining the displacement $\xi_o + y_o$, till the initial time $t_o=0$. Before $t_o$, the system is at rest and the velocity of any point is zero.

The displacement of the spring at $x=L$ and $t=0$:

$$y^* = 0 \quad \text{for } x = 0; \quad y^* = \xi_o \quad \text{for } x = L \tag{20}$$

To find the coefficients $A_\omega$, we impose at $t_o=0$:

$$\Sigma_\omega A_\omega \sin[\omega(\rho/E)^{1/2} x] = x\xi_o/L . \tag{21}$$

Coefficients $A_\omega$ can be deduced from (21) after the angular frequencies $\omega$ have been determined.
To find the set of frequencies, we study the motion of mass $M$. Note that the displacement of $M$ coincides with that of the end of the spring. The acceleration of $M$ is then $\partial^2 y(l, t)/\partial t^2$ and the equation of motion for $M$ is:

$$-SE(\partial y/\partial x)_{x=L} + Mg = M\partial^2 y(L, t)/\partial t^2 \tag{22}$$

After equations (16) and (19), we have:

$$S(\rho E)^{1/2} \omega \cos[\omega(\rho/E)^{1/2} L] = m \omega^2 \sin[\omega(\rho/E)^{1/2} L] \tag{23}$$

Using the dimensionless parameter:

$$p = \omega(\rho/E)^{1/2} L \tag{24}$$

and having $\rho S L = m$, Eq.(23) turns out to be:

$$p \sin p = (m/M) \cos p \tag{25}$$

This equation has infinite solutions $p_1, p_2, .... p_n, ...$ giving the set of angular frequencies:

$$\omega_n = L^{-1}(E/\rho)^{1/2} p_n \tag{26}$$

The general solution is:

$$y(x, t) = \Sigma_n A_n \sin(p_n x/L) \cos[L^{-1}(E/\rho)^{1/2} p_n t] + A'x + B'x^2 \tag{27}$$

where $A_n$ are coming from the condition

$$\Sigma_n A_n \sin(p_n x/L) = x\xi_0/L \tag{28}$$

The motion of the mass is the superposition of many harmonic motions, creating a motion which is *not* harmonic. However coefficients $A_n$, which can be obtained from (28), are decreasing in magnitude as *n* is increasing: a good approximation is then to consider only the coefficient $A_1$ being different from zero. This corresponding to the first root $p_1$ of (24), which is greater than 0 and less than $\pi/2$. With this approximation, the motion is harmonic, with period $\tau = 2\pi/\omega_1$, or:

$$\tau = 2\pi L (\rho/E)^{1/2} / p_1 \tag{29}$$

Figure 3 shows the behaviour of the first root *p*, as a function of the mass ratio *M/m*. Some values of the first root are given in the Table I.

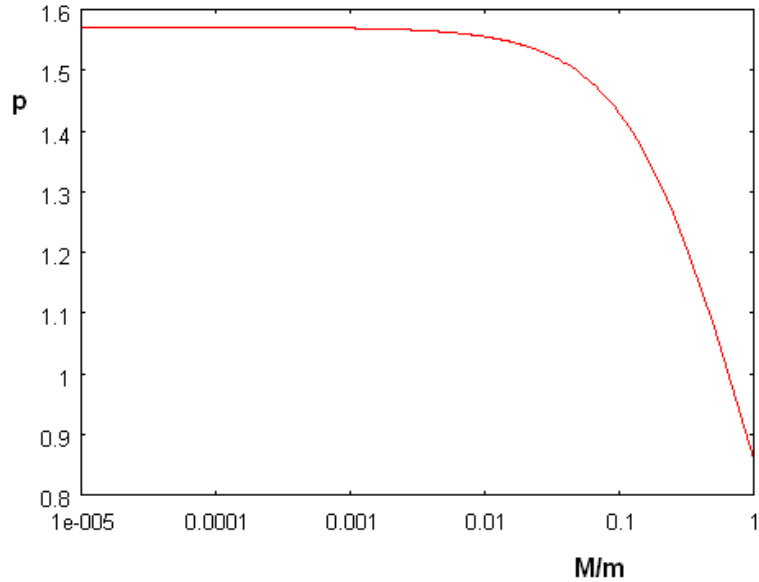

Figure3: Behaviour of first *p* as a function of the mass ratio *M/m*. The limit value for very small mass ratios is the value $\pi/2$ given by Eqs. 15,15'.

| M/m | $p_1$ | M/m | $p_1$ |
|---|---|---|---|
| 0.25 | 1.2646 | 1.75 | 0.6910 |
| 0.50 | 1.0769 | 2.0 | 0.6532 |
| 0.75 | 0.9512 | 2.25 | 0.6211 |
| 1.0 | 0.8603 | 2.50 | 0.5932 |
| 1.25 | 0.7910 | 2.75 | 0.5688 |
| 1.50 | 0.7360 | 3.0 | 0.5472 |

Table I: Values of first *p* for several mass ratios *M/m*.

Let us consider $\tau = 2\pi/\omega_1$, that is $\tau = \tau_0 / p_1$, $\tau_0 = 2\pi L (\rho/E)^{1/2}$. Period $\tau_0$ can be obtained from measurements, plotting as a function of the suspended mass *M*, the measured period $\tau$ times $p_1$. This product is a constant equal to $\tau_0$. Using masses $M_1, M_2, ... M_i, ... M_N$ and having $\tau_1, \tau_2, ... \tau_i ... \tau_N$, the value $\tau_0$ turns out to be:

$$\tau_0 = N^{-1} \Sigma_i \tau_i p_1(\mu_i) = N^{-1} \Sigma_i \tau_{0,i} \tag{30}$$

where $\mu_i = M_i/m$. The constant of the spring is $k = S E / L$, and then $k = (2\pi/\tau_0)^2 m$.
In this approach we avoided the use of an effective mass, which is determined in Ref.3.

**Discussion**

Let us report the results of experimental measurements obtained with a heavy spring of $m = 0.057$ kg, with several suspended masses, as given in the first column of Table II. The second column is reporting the corresponding value of $p_1$ as from Eq.25. $\tau$ is the measured period. Note that the product $p_1\tau$, is constant (under experimental uncertainty). The last column gives the force constant of the spring defined as $k_i = (2\pi/\tau_{0, i})^2 m$, in one measurement.

| m/M | $p_1$ from Eq.25 | $\tau$ (s) measured | $p_1 \tau$ (s) | $k_i$ (N/m) |
|---|---|---|---|---|
| 0.520 | 0.664 | 0.636 | 0.4226 | 12.601 |
| 0.562 | 0.686 | 0.609 | 0.4183 | 12.861 |
| 0.570 | 0.690 | 0.612 | 0.4228 | 12.589 |
| 0.449 | 0.624 | 0.674 | 0.4203 | 12.738 |
| 0.389 | 0.586 | 0.711 | 0.4169 | 12.947 |
| 0.297 | 0.519 | 0.806 | 0.4186 | 12.842 |

Table II

From Eq.30, the force constant of the heavy spring is 12.76 N/m. As uncertainty of the measurement, let us assume $\pm (12.95-12.59)/2$ N/m $= \pm 0.18$ N/m. The periods have been measured with stop-watches, over 20 oscillations, avoiding the count of the first oscillations.

We can compare this dynamical evaluation of the force constant with that obtained from static measurements. Let us use a reference mass $m_{ref} = 0.0648$ kg; this mass provokes a deformation, which adds to the deformation given by gravity. We measure a distance $d_{ref}$ of 16.0 cm, as that given in Figure 4. This is the distance from the supporting stand to the hook at the end of the spring (see Fig.4).

The force constant of the heavy spring will be:

$k = g (M - m_{ref})/(D - d_{ref})$ (31)

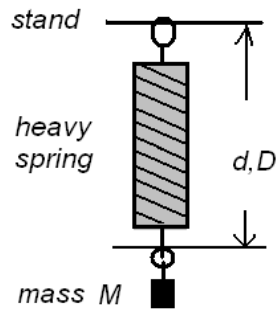

Figure 4

Table III shows the result of measurements, with several hanged masses, for $g=9.8$ m/s$^2$.

| M (kg) | D (cm) | k (N/m) |
|---|---|---|
| 0.0817 | 17.3 | 12.740 |
| 0.1096 | 19.4 | 12.913 |
| 0.1014 | 18.65 | 13.535 |
| 0.1000 | 18.5 | 13.798 |
| 0.1270 | 20.5 | 13.546 |
| 0.1918 | 25.5 | 13.101 |

Table III

The average value is 13.27 *N/m*. The uncertainty is ± (13.80−12.74)/2 *N/m* = ± 0.53 *N/m*. There is then a small overlap between the result of this static measurement with the previously proposed dynamical one.

The static evaluation is based on the measurement of lengths. In the laboratory, we used a rule for carpentry and this is why we proposed in Eq.31 to choose a reference mass and a reference length, measured with the spring supporting the reference mass, instead of the length of the heavy spring at rest under the effect of its sole weight. The rule for carpentry is not the best tool for measuring distances, but the proposed approach can be surely refined.